\documentclass[a4paper,12pt]{article}
\usepackage[pctex32]{graphics}
\usepackage{amssymb,amsmath}

\begin{document}

\topmargin 0pt
\oddsidemargin 0mm
\newcommand{\be}{\begin{equation}}
\newcommand{\ee}{\end{equation}}
\newcommand{\ba}{\begin{eqnarray}}
\newcommand{\ea}{\end{eqnarray}}
\newcommand{\fr}{\frac}

\renewcommand{\thefootnote}{\fnsymbol{footnote}}

\begin{titlepage}

\vspace{5mm}
\begin{center}
{\Large \bf Entropy function approach to charged BTZ black hole}

\vskip .6cm
 \centerline{\large
 Yun Soo Myung$^{1,a}$, Yong-Wan Kim $^{1,b}$
and Young-Jai Park$^{2,c}$}

\vskip .6cm

{$^{1}$Institute of Basic Science and School of Computer Aided
Science,
\\Inje University, Gimhae 621-749, Korea \\}

{$^{2}$Department of Physics and Center for Quantum Spacetime,\\
Sogang University, Seoul 121-742, Korea}
\end{center}

\vspace{5mm}
 \centerline{{\bf{Abstract}}}

\vspace{5mm}

We find solution to the metric function $f(r)=0$ of charged BTZ
black hole making use of the Lambert function. The condition of
extremal charged BTZ black hole is determined  by a non-linear
relation of $M_e(Q)=Q^2(1-\ln Q^2)$. Then, we study the entropy of
extremal charged BTZ black hole using the entropy function
approach. It is shown that this formalism works
with a proper normalization of charge $Q$ for charged BTZ
black hole because AdS$_2$$\times$S$^1$  represents near-horizon
geometry of the extremal charged BTZ black hole. Finally, we
introduce the Wald's Noether formalism to reproduce the  entropy
of the extremal charged BTZ black hole without normalization
when using the dilaton gravity approach.

\vspace{5mm}


\vskip 0.8cm

\noindent $^a$ysmyung@inje.ac.kr \\
\noindent $^b$ywkim65@gmail.com \\
\noindent$^c$yjpark@sogang.ac.kr

\noindent
\end{titlepage}

\newpage

\renewcommand{\thefootnote}{\arabic{footnote}}
\setcounter{footnote}{0} \setcounter{page}{2}

\section{Introduction}
Counting microstates using the AdS/CFT correspondence~\cite{Mald}
works well only when black hole geometry factorizes as AdS$_3
\times $M or AdS$_2 \times$M~\cite{SV}. Thus, AdS$_3$ and AdS$_2$
quantum gravity together with 3D and 2D black holes in AdS
spacetimes play an important role in computing the statistical
entropy of their black holes.  The  AdS$_3$ quantum gravity could
be identified with a dual 2D conformal field theory (CFT$_2$) with
the central charge $c=3G/2l$, which describes Brown-Henneaux
boundary excitations~\cite{BH}, that is, deformations of the
asymptotic boundary of AdS$_3$. This is possible because
asymptotic isometry group of AdS$_3$ is exactly conformal group of
CFT$_2$. Then, the CFT provides correctly the entropy of
Banados-Teitelboim-Zanelli (BTZ) black hole and a wide class of
higher-dimensional black holes when using the Cardy's
formula~\cite{Strom1}.

On the other hand, the AdS/CFT correspondence in two dimensions is
quite enigmatic~\cite{CM0,Strom2,CM1,Nav,CM2}. It is not clear
whether AdS$_2$ quantum gravity has to be regarded as either  the
chiral half of CFT$_2$ or conformal quantum mechanics (CFT$_1$) on
the asymptotic one-dimensional boundary of AdS$_2$. The first
version of AdS$_2$/CFT$_1$ correspondence, which was constructed
closely from the Brown-Henneaux formulation of AdS$_3$ quantum
gravity,  is based on AdS$_2$ endowed with a linear dilaton
background. Recently, the second version of AdS$_2$/chiral CFT$_2$
correspondence was proposed by considering a constant dilaton and
Maxwell field~\cite{HS} and its applications~\cite{AA}. A
circularly symmetric dimensional reduction allows us to describe
AdS$_3$ as AdS$_2$ with a linear dilaton. More recently, it has
been proposed that the charged BTZ black
hole~\cite{Clem1,Clem2,MTZ} may interpolate between two different
versions of AdS$_2$ quantum gravity, asymptotic AdS$_3$ and a
near-horizon AdS$_2 \times$S$^1$~\cite{CS,CMP}.

Generally, the AdS$_2$ quantum gravity could be used to derive the
entropy of extremal BTZ black hole when applying  the entropy
function formalism to the near-horizon geometry  factorized as
AdS$_2\times$M of extremal black holes~\cite{SSen,AFM,mkp1}. In
this case, the attractor equations work exactly as the Einstein
equations on AdS$_2$ do.

In this work, we  find solution to  the metric function $f(r)=0$
of charged BTZ black hole making use of the Lambert function. We
show that the entropy function formalism  works for charged BTZ
black hole even though the condition of extremal charged BTZ black
hole is special as given by a non-linear relation of
$M_e(Q)=Q^2(1-\ln Q^2)$, compared to others. It  suggests that
charged BTZ black hole may be a curious  ground for obtaining the
entropy of extremal black hole. Furthermore, we show that the
dilaton gravity approach reproduces the  entropy of extremal
charged BTZ black hole when using the Wald's Noether charge
formalism~\cite{Wald,Cai07,mkp4}.

\section{The charged BTZ black hole}
AdS$_3$ gravity admits the charged black hole solution when coupled
with the Maxwell term.
 The Martinez-Teitelboim-Zanelli action~\cite{MTZ} is given by
\begin{equation} \label{mtz}
 I_{MTZ}\equiv \int d^3x {\cal L}=\frac{1}{16\pi G}\int d^3x \sqrt{-g}
 \left[R+\frac{2}{l^2}-F_{mn}F^{mn}\right],
\end{equation}
where $F_{mn}$ is the electromagnetic field strength. The Latin
indices $m,n,\cdots$ represent three dimensional tensor. Equations
of motion for $A_m$ and $g_{mn}$ lead to
\begin{eqnarray}
 && \label{EOM-A}
  \partial_\nu\left(\sqrt{-g}F^{\mu\nu}\right)=0,\\
 && \label{EOM-g}
   R_{mn}-\frac{1}{2}g_{mn}R-\frac{1}{l^2}g_{mn}
   =2\left(F_{mp}F_n^{~p}-\frac{1}{4}g_{mn}F_{pq}F^{pq}\right).
\end{eqnarray}
The trace part of Eq. (\ref{EOM-g}) takes the form
\begin{equation}
 \label{trgEOM}
 R+\frac{6}{l^2}+F_{pq}F^{pq}=0.
\end{equation}
Here we have two parameter family $(M,Q)$ of electrically charged
black hole solutions
\begin{eqnarray}\label{metric}
ds^2&=&-f_w(r)dt^2+\frac{dr^2}{f_w(r)}+r^2d\theta^2, \\
f_w(r)&=&-M+\frac{r^2}{l^2}-Q^2
\ln\Big[\frac{r^2}{\omega^2}\Big],~~~F_{tr}= \frac{Q}{r},
\end{eqnarray}
where $M,\omega$ are constants and $-\infty<t<\infty,~0\le
r<\infty,~0\le \theta < 2\pi$. We also choose $G=1/8$ for the sake
of simplicity. A crucial difference with the BTZ black hole is the
presence of a power-law singularity $(R \sim 2Q^2/r^2$) at $r=0$
when one uses Eq. (\ref{trgEOM}). We note that the charged BTZ
black hole has two unpleasant features. Firstly, the mass $M$ is
not well defined because one gets logarithmic divergent boundary
terms when varying the action. That is, since the Maxwell
potential $A_t=-Q\ln(r)$ diverges logarithmically, the
 mass $M$ is ambiguously defined.
Secondly, it seems that the location of extremal charged BTZ black
hole is clearly determined from the condition of $f_w'(r_e)=0$ as
$r_e=lQ$ because both $M$ and the logarithmic function disappear
in $f_w'(r)$. However, it seems that  the near-horizon geometry
AdS$_2\times S^1$ of extremal charged BTZ black hole is not
uniquely defined because of $f_w''(r_e)=4/l^2$, which  shows that
the AdS$_2$-curvature $R_2=-f_w''(r_e)$ is independent of the
charge ``$Q$" but it depends on the cosmological constant. We may
regard this as a peculiar property of charged BTZ black hole.

\begin{figure}[t!]
   \centering
   \includegraphics{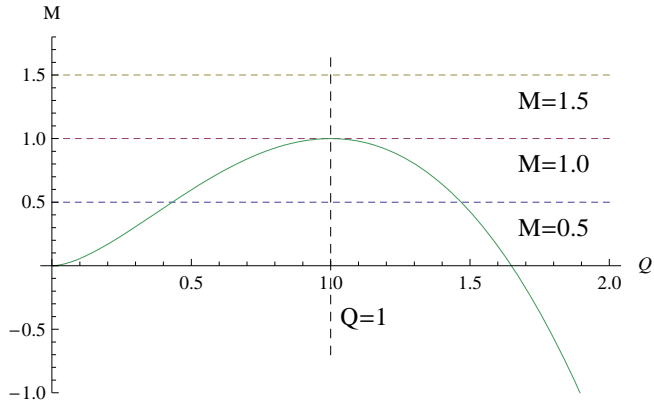}
\caption{Region of mass-charge plane with $l=1$. The curve
represents a non-linear relation $M_e(Q)=Q^2[1-\ln(Q^2)]$ for the
extremal charged BTZ black hole. The horizontal lines of $M=0.5$,
$M=1.0$, and $M=1.5$ are chosen to represent the characteristic of
the charged BTZ black hole. A part of the curve in $0\le Q<1$
could represent the known BTZ and Reissner-Nordstr\"om (RN) black
holes.} \label{fig.1}
\end{figure}

Furthermore, to avoid naked singularities, one imposes a BPS-like
bound for $M$ and $Q$ using the value of $-f_w$ at the minimum
\begin{equation}
\Delta \equiv M-Q^2\Big[1-\ln(Q^2)\Big].
\end{equation}
If $\Delta >0$, there are two zeros of $f_w(r)$; inner ($r_-$) and
outer ($r_+$) horizons. For $\Delta =0$, the two roots coincide
and it becomes the extremal black hole. At the extremal points of
$\Delta=0$, the mass is zero ($M_e=0$) at $Q=0$, has a maximum
($M_e=1$) at $Q=1$, vanishes ($M_e=0$) at $Q=\sqrt{e}$, and tends
to negative infinity ($M_e\to -\infty$) for large $Q$. This is
depicted in Fig. 1. The first problem may  be handled by
introducing a regularized metric function $f_r$~\cite{CMP,CMS}
\begin{equation} \label{f1}
f_r(r)\equiv
-M_0(r_0,\omega)+\frac{r^2}{l^2}-Q^2\ln\Big[\frac{r^2}{r^2_0}\Big],
~~M_0(r_0,\omega)=M+Q^2\ln\Big[\frac{r^2_0}{\omega^2}\Big].
\end{equation}
The parameter $\omega$ is considered as a running scale and
$M_0(r_0,\omega)$ is a regularized black hole mass, as  sum of
gravitational and electromagnetic energies inside a circle of
radius $r_0$. However, the second issue on near-horizon geometry
could not be resolved even if one chooses $f_r$, instead of $f_w$.
In this work, we are interested mainly in the near-horizon
geometry of the extremal charged  black hole. Hence, we use  the
metric function $f_w$ with $w=l$~\cite{Med,AWD,JWC,mkp3}
\begin{equation} \label{f2}
f_l\to f(r)=-M+\frac{r^2}{l^2}-Q^2\ln\Big[\frac{r^2}{l^2}\Big].
\end{equation}
It is well known that the charged BTZ black hole has the inner
($r=r_-$) and outer ($r=r_+$) event horizons which satisfy
$f(r_\mp)=0$. However, as far as we know, there is no explicit
forms of these horizons. The presence of the logarithmic term
makes it difficult to find explicit forms of two horizons.

\begin{figure}[t!]
   \centering
   \includegraphics{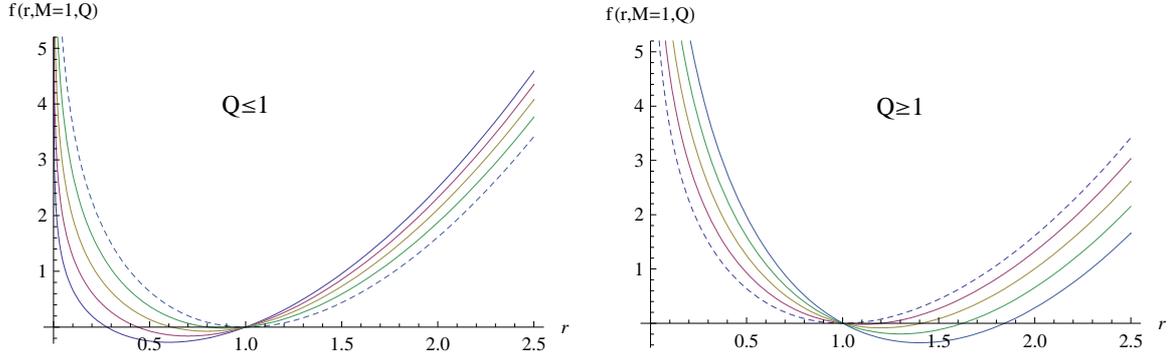}
\caption{Left panel:  metric functions $f(r,M=1,Q)$ for different
values $Q(\le1)=0.6, 0.7,0.8,0.9,1.0$ with $M=1$ and $l=1$ from
bottom to top. This shows clearly that as the charge increases,
the inner horizon $r_-$ increases while the outer horizon $r_+$
remains fixed.  Right panel:  metric functions $f(r,M=1,Q)$ for
different values $Q(\ge1)=1.0, 1.1, 1.2, 1.3, 1.4$  from bottom to
top. This shows that as the charge increases, the outer horizon
$r_+$ increases while the inner horizon $r_-$ remains fixed. The
dotted curves are for the extremal black holes. } \label{fig.2}
\end{figure}

By introducing  new coordinates $\tau$ and $\rho$ in Eqs.
(\ref{metric}) and (\ref{f2}) as
\begin{equation}
\tau=\frac{2\epsilon}{l^2}t,~~~\rho=\frac{r-Ql}{\epsilon},
\end{equation}
Eq. (\ref{metric}) leads to  the near-horizon geometry of extremal
charged BTZ black hole,  AdS$_2$$\times$S$^1$ in the limit of
$\epsilon \to 0$
\begin{equation}\label{ads}
ds^2_{NHEB}=v_1(-\rho^2d\tau^2+\frac{1}{\rho^2}d\rho^2)+v^2_2d\theta^2,
\end{equation}
with \be \label{spads}
v_1=\frac{f''(r_e)}{2}=\frac{l^2}{2},~~v_2=Ql.\ee It seems that
Eq. (\ref{ads}) represents the near-horizon geometry of the
extremal black hole. However, we observe that the
AdS$_2$-curvature radius $v_1$ does not depend on the charge
``$Q$". The disappearance of the charge is mainly due to the
logarithmic function of $-Q^2\ln [r^2/l^2]$ in $f(r)$: its first
derivative is $-2Q^2/r$ and the second derivative takes the form
$2Q^2/r_e^2=2/l^2$ at $r=r_e$. Hence, it is shown that the origin
of the disappearance of the charge is because we consider the
``charged" BTZ black hole in three dimensions.

 Qualitatively, one can further analyze the metric function
(\ref{f2}) to see the outer/inner horizon behaviors according to
values of the mass and charge. Firstly, for $M=1$, as is shown in
Fig. 2, there are two opposite cases according to the values of
the charge. In the left panel of Fig. 2, for $M=1$ and $Q< 1$, the
inner horizon $r_-$ increases as the charge $Q$ increases, while
the outer horizon $r_+$ remains fixed. On the other hand, in the
right panel of the Fig. 2, we find that for $Q> 1$, the outer
horizon $r_+$ increases as $Q$ increases, while the inner horizon
$r_-$ remains fixed. On the other hand, for $M=Q=1$, two horizons
coincide and it becomes extremal black hole as shown in Fig.2.

Secondly, for $M=0.5$ between $0<M<1$, as shown in the left panel
of Fig. 3, the inner horizon $r_-$ increases while the outer
horizon $r_+$ decreases as the charge $Q$ increases. On the other
hand, the right panel of Fig. 3 shows that as $Q$ increases the
outer horizon $r_+$ increases, while the inner horizon $r_-$
decreases. Note that for very small Q in regions of
$0<Q<Q_e=0.432$, the outer horizon approaches a constant value
($r_+\rightarrow 1$), while for large Q in regions of
$Q>Q_e=1.467$, the inner horizon approaches a constant value
($r_-\rightarrow 1$).
\begin{figure}[t!]
   \centering
   \includegraphics{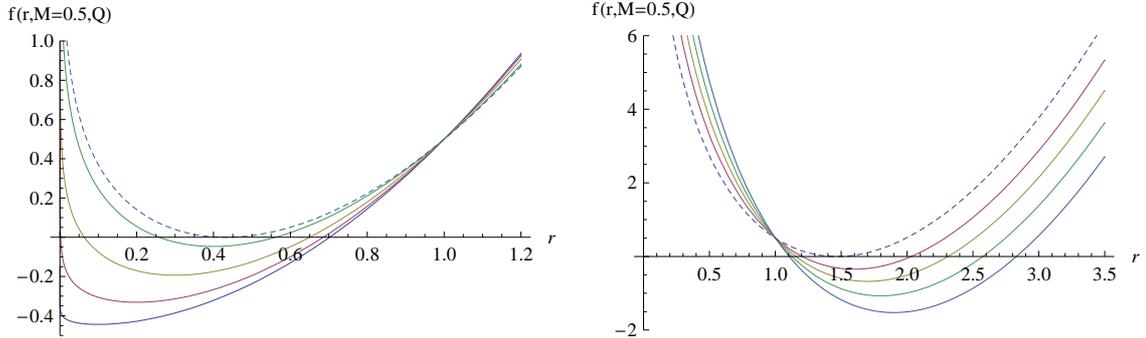}
\caption{Left panel:  metric functions $f(r,M=0.5,Q)$ for
different values $Q(Q_e\le 0.432)=0.1, 0.2, 0.3, 0.4, 0.432$ with
$M=0.5$ and $l=1$ from bottom to top. As the charge increases, the
inner horizon $r_-$ increases while the outer horizon $r_+$
decreases.  Right panel:  metric functions $f(r,M=0.5,Q)$ for
different values $Q(Q_e\ge1.467)=1.467, 1.6, 1.7, 1.8, 1.9$ from
top to bottom for curves in $r>1$. This shows that as the charge
increases, the outer horizon  $r_+$ increases while the inner
horizon $r_-$ decreases. } \label{fig.3}
\end{figure}
\begin{figure}[t!]
   \centering
   \includegraphics{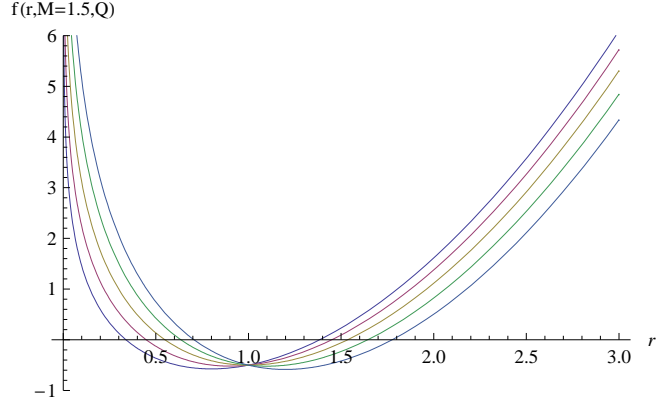}
\caption{Metric functions $f(r,M=1.5,Q)$ for different values $Q=
0.8, 0.9, 1.0, 1.1, 1.2$ with  $M=1.5$ and $l=1$ from bottom to
top for curves in $0<r<1$. As the charge increases, the inner
horizon $r_-$ increases, and the outer horizon $r_+$ also
increases. This implies that there is no extremal black hole. }
\label{fig.4}
\end{figure}

Thirdly, for $M>1$, as shown in Fig. 4, both the inner horizon
$r_-$ and the outer horizon $r_+$ increase as the charge $Q$
increases. In this case, there is no extremal black hole as
expected, and for large $Q$ the inner horizon approaches
$r_-\rightarrow 1$. We will check these qualitative behaviors of
the metric function by solving $f(r_\mp)=0$ explicitly.

\begin{figure}[t!]
   \centering
   \includegraphics{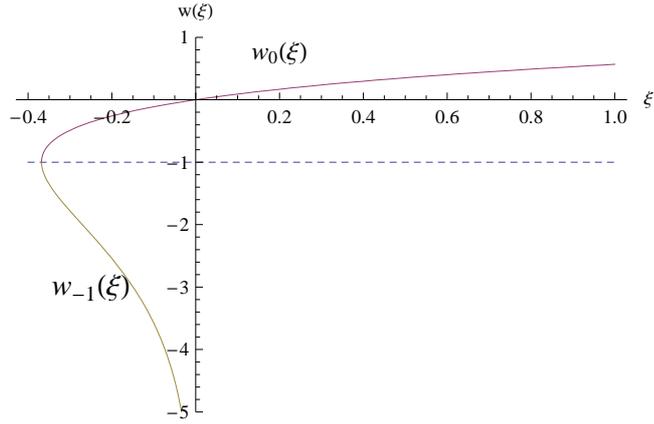}
\caption{Lambert functions $w_0(\xi)$ and $w_{-1}(\xi)$.
$w_0(\xi)$ represents the upper branch ($w_0(\xi)\ge -1$) for
$\xi\ge -\frac{1}{e}$, while the lower branch denotes
$w_{-1}(\xi)\le -1$ for $-\frac{1}{e}\le \xi \le 0$. At
$\xi=-\frac{1}{e}=-0.368$, one finds $w_0=w_{-1}$ for the extremal
charged BTZ black hole. } \label{fig.5}
\end{figure}

\section{Exact solution to $f(r)=0$}

Now let us find the exact solution of $f(r_\mp)=0$ by using the
Lambert functions $w_k(\xi)$. As is shown in Fig. 5, $w_0(\xi)$
and $w_{-1}(\xi)$ are two real functions~\cite{Maty,mkp2}. With
$1/x=\ln(r/l)$, $f(r)=0$ takes the form
\begin{equation}
x\Big(e^{\frac{2}{x}}-M\Big)=2Q^2.
\end{equation}
In order to solve this, we introduce $2/x=a w(\xi)+b $ with $a$
and $b$ two unknown constants. The above equation leads to
\begin{equation}
e^{-a w} aw=\frac{1}{Q^2} e^b,~~b=-\frac{M}{Q^2}.
\end{equation}
Choosing $a=-1$,  one has  the following equation to define the
Lambert function $w(\xi)$
\begin{equation}
e^w(\xi) w(\xi)=\xi
\end{equation}
with
\begin{equation}
\xi=-\frac{1}{Q^2} e^{-\frac{M}{Q^2}}.
\end{equation}
Then, two horizons are determined  by
\begin{eqnarray}
 r_-(M,Q)&=&l
 \exp\left[-\frac{Q^2w_0\Big(-\frac{1}{Q^2}e^{-\frac{M}{Q^2}}\Big)+M}{2Q^2}\right],\nonumber\\
 ~~r_+(M,Q)&=&l\exp\left[-\frac{Q^2w_{-1}\Big(-\frac{1}{Q^2}e^{-\frac{M}{Q^2}}\Big)+M}{2Q^2}\right].
\end{eqnarray}
\begin{figure}[t!]
   \centering
   \includegraphics{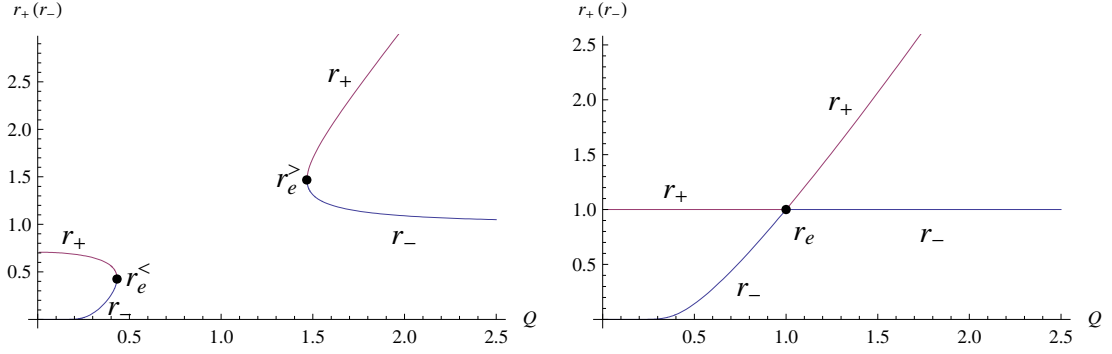}
\caption{Left panel: two horizons  $r_+$ and $r_-$ as functions of
$Q$ with  $M=0.5(<1)$. For $Q<Q_e (Q>Q_e)$, the outer horizon
$r_+$ decreases (increases), while the inner horizon $r_-$
increases (decreases and approaches 1). Right panel: two horizons
$r_+$ and $r_-$ for $M=1$. For $Q<1 (Q>1)$, the inner horizon
$r_-$ (the outer horizon $r_+$) increases, while the outer horizon
$r_+$ (the inner horizon $r_-$) remains fixed. The extremal point
of $r_e$ indicates the location of extremal black hole. }
\label{fig.6}
\end{figure}
We check that the extremal black hole appears when $r_-=r_+=r_e$.
In this case, we have $w_0=w_{-1}=-1$ at $\xi=-1/e$ so that
\begin{equation}
r_e=lQ,~M_e=Q^2\Big[1-\ln(Q^2)\Big].
\end{equation}
The left panel of Fig. 6 shows that for $M=0.5$ between $0<M<1$
with $Q<1$, the outer horizon $r_+$ is a monotonically decreasing
while the inner horizon $r_-$ is a increasing function of $Q$. Two
horizons coincide at $r_+=r_-=r_e^<$ to be an extremal black hole
at $Q=Q_e$. On the other hand, for $Q>1$, the outer horizon $r_+$
is monotonically increasing while the inner horizon $r_-$
approaches a constant value. It confirms  the qualitative results
in the Fig. 3. Note that in the case of $0<M<1$ with $Q<1$, the
behavior of two horizons is the nearly same with that of the BTZ
black holes~\cite{BTZ} (see also Fig. 7.) whose horizons are given
by
\begin{equation} \label{btzmet}
r^{BTZ}_{\pm}=l \sqrt{\frac{M}{2}} \Bigg\{1 \pm
\Bigg[1-\Bigg(\frac{J}{Ml}\Bigg)^2\Bigg]^{\frac{1}{2}}\Bigg\}^{\frac{1}{2}}.
\end{equation}
These horizons exist provided $(Ml)^2 \ge J^2$ and coalesce if
$Ml=J$ (the extremal case). As is shown in Fig. 7,
$r^{BTZ}_\pm(M=1,J)$ for the BTZ black hole and
$r^{RN}_\pm(M=1,Q)$ for the RN black hole are the nearly same with
that of the charged BTZ black hole for $M=0.5$ between $0<M<1$
with $Q<1$. As $J$ increases in the BTZ black hole, the outer
horizon $r_+$ decreases, while the inner horizon $r_-$ increases.
Similarly, as $Q$ increases in the RN black hole, the outer
horizon $r_+$ decreases, while the inner horizon $r_-$ increases.
Both cases imply that other branches of $Ml <J$ and $M <Q$ are not
allowed.

\begin{figure}[t!]
   \centering
   \includegraphics{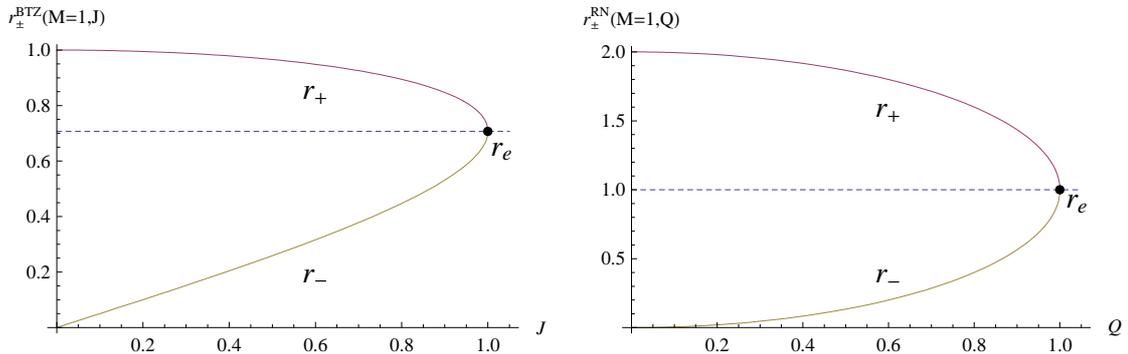}
\caption{Left panel: two horizons  $r_+$ and $r_-$ as functions of
$J$ with  $M=1$ for the BTZ black hole in Eq.(\ref{btzmet}) with
$l=1$.
 We confirm that $r_+=r_-=\frac{1}{\sqrt{2}}$ at
$J=1$ indicates the location of extremal BTZ black hole. Right
panel: two horizons $r^{RN}_\pm=M\pm\sqrt{M^2-Q^2}$ as functions
of $Q$ with  $M=1$ for the RN black hole. We confirm that
$r_+=r_-=1$ at $Q=1$ denotes  the location of extremal RN black
hole. } \label{fig.7}
\end{figure}

\begin{figure}[t!]
   \centering
   \includegraphics{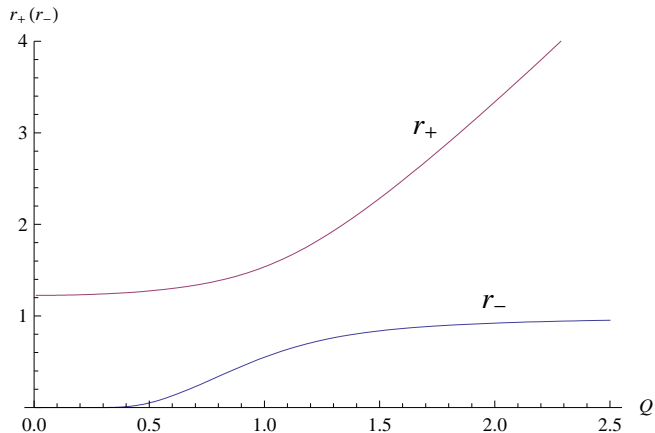}
\caption{Two horizons $r_+$ and $r_-$ for $M=1.5(>1)$. There is no
extremal black hole, the outer horizon $r_+$ is a increasing
function of $Q$ , while the inner horizon $r_-$ approaches  a
constant value as $Q$ increases. } \label{fig.8}
\end{figure}

On the other hand, the behavior of the horizons in the right-hand
side of the left panel of Fig. 6 shows that the charged BTZ black
hole is basically  different from  BTZ and RN black holes. As the
charge $Q$ increases, the outer horizon $r_+$ is increasing while
the inner horizon $r_-$  approaches a constant value. Moreover,
the left panel of Fig. 6 shows that there is a forbidden region
between the small (left) and large (right) extremal charges for
$M=0.5$ in $0<M<1$. For $M=M_e=1$, the  right-panel of Fig. 6
shows no such forbidden region and as the charge increases,  the
outer horizon $r_+$ (the inner horizon $r_-$) is fixed  while the
inner horizon $r_-$ (the outer horizon $r_+$) increases.  This
confirms that the numerical results in  Figs. 2 and 3 are correct.
Finally,  Fig. 8 shows that for $M=1.5(M>1)$, there is no extremal
black hole as expected and the outer horizon $r_+$ is a
monotonically increasing function of $Q$ while the inner horizon
$r_-$ approaches a constant value.

Up to now, we show that the extremal charged BTZ black hole
depends heavily on the charge ``$Q$" as well as the mass ``$M$".

Finally, the Bekenstein-Hawking entropy for the extremal charged
BTZ black hole is defined by
\begin{equation} \label{BHE}
S_{BH}=\frac{\pi r_e}{2G}= 4 \pi Q l
\end{equation}
with $G=1/8$.

\section{Solution to Einstein equations on  AdS$_2\times S^1$ }
Since the near-horizon geometry (\ref{ads}) with (\ref{spads}) of
extremal charged BTZ black hole is different from those of BTZ and
RN black holes, it is very interesting to find the entropy of its
black hole. In order to obtain the entropy of extremal charged BTZ
black hole, we assume the near-horizon geometry AdS$_2\times$S$^1$
of the extremal charged BTZ black hole as
\begin{equation}
ds^2=v_1\left(-r^2dt^2+\frac{dr^2}{r^2}\right)+v^2_2d^2\theta
\end{equation}
with
\begin{equation} \label{ads2inf}
 R=-\frac{2}{v_1}, ~~~F_{mn}F^{mn}=-\frac{2e^2}{v^2_1},
 ~~~F_{01}=e,~~~\sqrt{-g}=v_1v_2.
\end{equation}
We wish to solve the Einstein equations (\ref{EOM-A}),
(\ref{EOM-g}), and (\ref{trgEOM}) on the AdS$_2\times S^1$. On the
AdS$_2$-background, Eq. (\ref{EOM-A}) is trivially satisfied,
while  $(tt)$- and $(rr)$-components of Eq. (\ref{EOM-g}) give as
\begin{eqnarray}
 \label{EE00}  &&  \frac{1}{l^2}=\frac{e^2}{v^2_1}, \\
 \label{EE22} && \frac{1}{v_1}-\frac{1}{l^2}=\frac{e^2}{v_1^2},
\end{eqnarray}
respectively. Note that  $(rr)$-component is duplicate because it
reproduces $(tt)$-one. Solving these, one obtains
\begin{equation}
\label{solEE} v_1=el=\frac{l^2}{2}.
\end{equation}
On the other hand, we could not determine ``$v_2$". We note that
the trace part (\ref{trgEOM}) of the Einstein equation is also
trivially satisfied upon using the solution (\ref{solEE}). In the
next section, we will check  these results by employing the
entropy function approach.

\section{Entropy function approach}

The entropy function \cite{SSen} is defined as the Legendre
transformation of ${\cal F}(v_1,v_2,e)$
\begin{eqnarray}
{\cal E}(v_1,v_2,q)=2\pi\left[qe-{\cal F}(v_1,v_2,e)\right],
\end{eqnarray}
where ${\cal F}(v_1,v_2,e)$ is obtained  by plugging Eq.
(\ref{ads2inf}) to ${\cal L}$ in Eq. (\ref{mtz})
\begin{equation}
 {\cal F}(v_1,v_2,e)={\cal L}(v_1,v_2,e)
 =2v_2\left(-1+\frac{v_1}{l^2}+\frac{e^2}{v_1}\right).
\end{equation}
Here we used $G=1/8$ after integration over ``$\theta$", and $q$
is a conserved quantity related to the charge $Q$ of the charged
BTZ black hole. Then, equations of motion for $v_1$, $v_2$, and
$e$ are given by (\ref{EE00}), (\ref{EE22}),  and
\begin{equation}\label{qv2}
    q=\frac{4ev_2}{ v_1},
\end{equation}
respectively. As a result, in addition to (\ref{solEE}),  the
solution is obtained as
\begin{equation}
v_2=\frac{ql}{4}.
\end{equation}
Plugging these into ${\cal E}$ leads to the  entropy function
\begin{equation} \label{entrof1}
{\cal E}\mid_{ext}=2\pi qe=\pi ql
\end{equation}
with a trivial identity of ${\cal F}\mid_{ext}=0$ due to Eq.
(\ref{EE22}). Note that ``$v_2$" is determined by the black hole
charge  $q$. Since $v_2$ determines the size of S$^1$ at the
horizon, one can use it to establish the relation between $q$ and
the charge $Q$ of the black hole, which is $q=4Q$. In the entropy
function approach, $q$ is always related to the charge ``$Q$" of
the charged BTZ black hole and thus, one may choose an appropriate
normalization ``4" to compare it with the Bekenstein entropy or
Wald's entropy.

 Now, according
to Eq. (\ref{entrof1}), one finds
\be \label{entrof2}
{\cal E}|_{ext} = 2 \pi qe = \pi ql = 4\pi Ql,
\ee
 which is in
perfect agreement with Eq. (\ref{BHE}). That is, one indeed gets
the correct entropy from the entropy function approach even though
${\cal F}\mid_{ext}=0$ is found.

\section{2D Maxwell-dilaton gravity}
In order to reconform the previous result (\ref{entrof2}),
let us use another approach, which is the Kaluza-Klein
dimensional reduction by considering the metric ansatz of
${\cal M}_2\times$S$^1$~\cite{AO,LMK,GM}
\begin{equation}
dS^2_{KK}=g_{\mu\nu}dx^\mu dx^\nu+\phi^2d^2\theta,
\end{equation}
where $\phi$ is the dilaton parameterizing the radius of the
$S^1$-sphere. Here the Greek indices $\mu,\nu,\cdots,$ represent
two-dimensional tensor. After the dimensional reduction, the 2D
Maxwell-dilaton action takes the form
\begin{equation} \label{2dd1}
 I_{2D}\equiv \int d^2x {\cal L}_2=\int d^2x \sqrt{-g}\phi
 \left[R+\frac{2}{l^2}-F_{\mu\nu}F^{\mu\nu}\right],
\end{equation}
where we choose $G=1/8$ for simplicity.
 Equations of motion for $\phi$, $A_\mu$ and
$g_{\mu\nu}$ are given by~\cite{CS}
\begin{eqnarray}
 && \label{EOM2-phi}
  R+\frac{2}{l^2}-F_{\mu\nu}F^{\mu\nu}=0,\\
 && \label{EOM2-A}
  \partial_\nu\left(\sqrt{-g}\phi F^{\mu\nu}\right)=0,\\
 && \label{EOM2-g}
  -\nabla_\mu\nabla_\nu\phi+\Big( \nabla^2\phi-\frac{\phi}{l^2}+\frac{\phi}{2}
  F_{\rho\sigma}F^{\rho\sigma} \Big)g_{\mu\nu}
   -2\phi F_{\mu\rho}F_\nu^{~\rho}=0,
\end{eqnarray}
respectively. It is important to note that these field equations
are invariant under rescaling of the dilaton like
$\tilde{\phi}={\cal C}\phi$ with an arbitrary constant ${\cal C}$.
Therefore, a constant mode of the dilaton may not be fixed. On the
other hand, the trace part of Eq. (\ref{EOM2-g}) leads to the
dilaton equation
\begin{equation}
 \label{trgEOM2}
 \nabla^2\phi-\frac{2\phi}{l^2}-\phi F_{\mu\nu}F^{\mu\nu}=0,
\end{equation}
and the traceless part of Eq. (\ref{EOM2-g}) takes the form
\begin{eqnarray}
 \label{trlessgEOM2}
 &&-\nabla_{\mu}\nabla_{\nu}\phi+ \frac{1}{2}g_{\nu\nu}\nabla^2\phi
    -2\phi(F_{\mu\rho}F_\nu^{~\rho} -\frac{1}{2} g_{\mu\nu}F_{\rho\sigma}F^{\rho\sigma})
    =0. \nonumber\\
\end{eqnarray}

Now, let us introduce the AdS$_2$ ansatz
\begin{equation}
ds^2=v\left(-r^2dt^2+\frac{dr^2}{r^2}\right)
\end{equation}
with
\begin{equation} \label{sol3}
R=-\frac{2}{v}, ~~~\phi=u,~~~
F_{\mu\nu}F^{\mu\nu}=-\frac{2e^2}{v^2},~~~F_{tr}=e,~~~\sqrt{-g}=v,
\end{equation}
which correspond to a constant dilaton and constant electric
field. The entropy function is defined as
\begin{eqnarray}
{\cal E}(u,v,q)=2\pi\left[qe-{\cal F}(u,v,e)\right],
\end{eqnarray}
where
\begin{equation}
 {\cal F}(u,v,e)={\cal L}(v,u,e)=2u\left(-1+\frac{v}{l^2}+\frac{e^2}{v}\right).
\end{equation}
Then, equations of motion for $v$, $u$, and $e$ are given by
\begin{eqnarray}
&& \label{eom2-v}
   \frac{1}{l^2}=\frac{e^2}{v^2},\\
 && \label{eom2-u}
    \frac{1}{v}-\frac{1}{l^2}=\frac{e^2}{v^2},\\
 && \label{eom2-e}
    q=\frac{4ue}{v},
\end{eqnarray}
respectively. As a result, solution is obtained as
\begin{equation} \label{sol4}
v=\frac{l^2}{2},~~u=\frac{ql}{4}.
\end{equation}
Plugging these into ${\cal E}$ leads to
 the entropy
\begin{equation} \label{entropy2}
{\cal E}\mid_{ext}=2\pi qe=\pi ql
\end{equation}
with the identity of ${\cal F}\mid_{ext}=0$ and undetermined
constant $u$. Similarly, we could determine the entropy
(\ref{BHE}) of the extremal charged BTZ black hole when choosing
an appropriate normalization $q=4Q$.

\section{2D dilaton gravity and Wald formalism}
In this section, we wish to find another method
to obtain the entropy in Eq. (\ref{BHE}) without normalization.

Let us derive an effective 2D dilaton gravity action by
integrating out the Maxwell field. Then, relevant fields will just
be the dilaton $\phi$ and metric tensor $g_{\mu\nu}$. First of
all, we solve Eq. (\ref{EOM2-A}) to have
\begin{equation}
\sqrt{-g}\phi F^{tr}= \tilde{Q},
\end{equation}
where $\tilde{Q}$ is an integration constant  related to the
charge $Q$  of the charged BTZ black hole. Considering the
 metric ansatz of $g_{\mu\nu}={\rm diag}
\{-f,f^{-1}\}$, $F_{tr}$ takes the form
\begin{equation}
F_{tr}=-\frac{\tilde{Q}}{\phi},
\end{equation}
which allows to express $F_{\mu\nu}^2$ as a function of the
dilaton $\phi$ with $\tilde{Q}$
\begin{equation}
F_{\mu\nu}F^{\mu\nu}=-\frac{2\tilde{Q}^2}{\phi^2}.
\end{equation}
We rewrite Eq. (\ref{trgEOM2}) as the dilaton equation
\begin{equation} \label{dd1}
\nabla^2\phi-V(\phi)=0
 \end{equation}
 with the dilaton potential parameterizing the original 3D theory
\begin{equation} \label{dipot}
V(\phi)=\frac{2\phi}{l^2}+\phi F_{\mu\nu}F^{\mu\nu}
=\frac{2\phi}{l^2}-\frac{2\tilde{Q}^2}{\phi}.
\end{equation}
Moreover, we can rewrite Eq. (\ref{EOM2-phi}) as the 2D curvature
equation
\begin{equation}
R+V'(\phi)=0 \label{dd2}
 \end{equation}
with
\begin{equation}
V'(\phi)=\frac{2}{l^2}+\frac{2\tilde{Q}^2}{\phi^2},
\end{equation}
where $'$ denotes the derivative with respect to $\phi$.
Importantly, we mention that two equations (\ref{dd1}) and
(\ref{dd2}) correspond to  attractor equations in the new
attractor mechanism~\cite{mkp4}. Actually, these equations could
be derived from the 2D dilaton action
\begin{equation} \label{2dd2}
I^{dil}=\int d^2x \sqrt{-g}
 \left[\phi R+V(\phi)\right].
\end{equation}
We note that the 2D Maxwell-dilaton  action (\ref{2dd1}) differs
from the 2D dilaton action (\ref{2dd2}), showing the sign change in
the front of the Maxwell term through Eq. (\ref{dipot}).

It is well known that $V(\phi)=0~(f'(r)=0)$ determines the
degenerate horizon for the extremal charged BTZ black hole with
$\tilde{Q}\equiv Q$ as \cite{mkp3}
\begin{equation}
\phi_e=Ql.
\end{equation}
Inserting this into the action (\ref{2dd2}) leads to
\begin{equation}
I^{dil}\mid_{ext}=\int d^2x {\cal F}^{dil}\mid_{ext}
\end{equation}
with Lagrangian density
\begin{equation}
{\cal F}^{dil}\mid_{ext}= R_e \phi_e,
\end{equation}
where
\begin{equation}
R_e=-V'(\phi_e)=-\frac{4}{l^2}=-\frac{2}{v}.
\end{equation}
Here the last equality confirms from Eqs. (\ref{sol3}) and
(\ref{sol4}).
 Using the Wald formula~\cite{Wald,Cai07,mkp4}, we obtain the entropy of the
 extremal charged BTZ black hole
\begin{equation}
  \label{phient}
  S=\frac{4\pi}{R_e}
\Big[{\cal F}^{dil}\mid_{ext}\Big]= 4 \pi \phi_e=4 \pi Q l,
\end{equation}
which  reproduces the Bekenstein-Hawking entropy in Eq.
(\ref{BHE}).

It was shown that the charged BTZ black hole solution could be
recovered exactly from its 2D dilaton gravity  with ``linear
dilaton $\phi=r$" when choosing $f(\phi)=-M+J(\phi)$
with~\cite{mkp3}
\begin{equation}
J(\phi)=\int^\phi_lV(\tilde{\phi})d\tilde{\phi}=\frac{\phi^2}{l^2}-Q^2\ln\Big[\frac{\phi^2}{l^2}\Big].
\end{equation}
Also its thermodynamic quantities of Hawking temperature $T_H$,
heat capacity $C$, and free energy $F$ are reproduced from the 2D
dilaton gravity of $V,V',J$ as
\begin{equation}
T_H=\frac{V(\phi)}{4\pi},~~C=4\pi \frac{V(\phi)}{V'(\phi)},
F=J(\phi)-J(\phi_e)-\phi V(\phi).
\end{equation}
We also confirm the condition of the extremal charged black hole:
$T_H(\phi_e)=0,~C(\phi_e)=0, F(\phi_e)=0$, in addition to the
entropy (\ref{phient}).

\section{Discussions}
Two different realizations of AdS$_2$ gravity show distinct
states.  AdS$_2$ quantum gravity with a linear dilation describes
Brown-Henneaux-like boundary excitations, which is suitable for
explaining the entropy of the charged BTZ black hole. On the other
hand, AdS$_2$ quantum gravity with a constant dilaton and Maxwell
field  may describe the near-horizon geometry of  the extremal
charged BTZ black hole.

As was shown in Figs. 2, 3, 4, 6, and 7, the charged BTZ black
hole with two horizons is  determined   by  the mass ``$M$" and
the charge ``$Q$". However, as Eqs. (\ref{ads}) and (\ref{spads})
are shown, its near-horizon geometry of the extremal charged black
hole is not uniquely determined by the charge ``$Q$". This is
compared to those for the extremal BTZ black hole and the extremal
RN black hole.

 In order to obtain the  entropy of
the extremal charged BTZ black hole, we use the entropy function
approach from the gravitational side. At this stage, we remind the
reader  the entropy function approach, which is based on the fact
that the near-horizon geometry depends on the charge $Q$, and it
is completely decoupled from the mass $M$, which is properly
defined at infinity. Hence one may conjecture that the charged BTZ
black hole is not a good model to derive its entropy using the
entropy function approach because the AdS radius does not depend
on the charge.

However,  we have shown that the entropy function formalism works
for obtaining the entropy of the extremal charged BTZ black hole.
 We check it by three different methods,
solving the Einstein equation on the AdS$_2$$\times$S$^1$, entropy
function, and 2D Maxwell-dilaton gravity approaches. This suggests
that the charged BTZ black hole may  be  a peculiar model to
obtain  the entropy of its extremal black hole when using the
entropy function formalism. On the other hand, the dilaton gravity
approach based on AdS$_2$ quantum gravity with a linear dilation
reproduces the correct entropy of the extremal charged BTZ black
hole when using the Wald's Noether formalism.

Consequently, the extremal  charged BTZ black hole was shown to
have a peculiar feature, in comparison with extremal BTZ and RN
black holes. We have  recovered the Bekenstein-Hawking entropy in
Eq. (\ref{BHE}) with an appropriate normalization $q=4Q$.

\section*{Acknowledgement}
Two of us (Y. S. Myung and Y.-J. Park) were supported by the
National Research Foundation of Korea (NRF) grant funded by the
Korea government (MEST) through the Center for Quantum Spacetime
(CQUeST) of Sogang University with grant number 2005-0049409. Y.-W.
Kim was supported by the Korea Research Foundation Grant funded by
Korea Government (MOEHRD): KRF-2007-359-C00007.


\begin{thebibliography}{99}
\bibitem{Mald}
  J.~M.~Maldacena,
  Adv.\ Theor.\ Math.\ Phys.\  2 (1998) 231
  [Int.\ J.\ Theor.\ Phys.\  38 (1999) 1113].
\bibitem{SV}
  A.~Strominger, C.~Vafa,
  Phys.\ Lett.\  B 379 (1996) 99.




\bibitem{BH}
  J.~D.~Brown, M.~Henneaux,
  Commun.\ Math.\ Phys.\  104 (1986) 207.

\bibitem{Strom1}
  A.~Strominger,
  JHEP 9802 (1998) 009.


\bibitem{CM0}
  M.~Cadoni, S.~Mignemi,
  Phys.\ Rev.\  D 59 (1999) 081501.



\bibitem{Strom2}
  A.~Strominger,
  JHEP 9901 (1999) 007.
\bibitem{MMS}
  J.~M.~Maldacena, J.~Michelson, A.~Strominger,
  JHEP 9902 (1999) 011.




\bibitem{CM1}
  M.~Cadoni, S.~Mignemi,
  Nucl.\ Phys.\  B 557 (1999) 165.


 \bibitem{Nav}
 J.~Navarro-Salas, P.~Navarro,
  Nucl.\ Phys.\  B  579 (2000)  250.

  \bibitem{CM2}
  M.~Cadoni, S.~Mignemi,
  Phys.\ Lett.\  B  490 (2000)  131.


\bibitem{HS}
T.~Hartman, A.~Strominger,
  JHEP 0904 (2009) 026.

\bibitem{AA}
  M.~Alishahiha, F.~Ardalan,
  JHEP  0808 (2008)  079.

  \bibitem{Clem1}
  G.~Clement,
  Phys.\ Lett.\  B  367 (1996)  70.

\bibitem{Clem2}
  G.~Clement,
  Class.\ Quant.\ Grav.\    10 (1993)  L49.



\bibitem{MTZ}
  C.~Martinez, C.~Teitelboim, J.~Zanelli,
  Phys.\ Rev.\  D  61 (2000)  104013.

\bibitem{CS}
M.~Cadoni, M.~R.~Setare,
  JHEP  0807 (2008)  131.

\bibitem{CMP}
M.~Cadoni, M.~Melis, P.~Pani,
  arXiv:0812.3362.

\bibitem{SSen}
  B.~Sahoo, A.~Sen,
  JHEP  0607 (2006)  008.


\bibitem{AFM}
  M.~Alishahiha, R.~Fareghbal, A.~E.~Mosaffa,
  JHEP  0901 (2009)  069.


\bibitem{mkp1}
 Y.~S.~Myung, Y.~W.~Kim, Y.-J.~Park,
 JHEP 0906 (2009) 043.

\bibitem{Wald}
R.~M.~Wald,
  Phys.\ Rev.\  D 48 (1993) 3427.

\bibitem{Cai07}
  R.~G.~Cai, L.~M.~Cao,
  Phys.\ Rev.\  D  76  (2007) 064010.

\bibitem{mkp4}
  Y.~S.~Myung, Y.~W.~Kim, Y.~J.~Park,
  Phys.\ Rev.\  D  76 (2007)  104045.

  \bibitem{CMS}
  M.~Cadoni, M.~Melis, M.~R.~Setare,
  Class.\ Quant.\ Grav.\  25 (2008)  195022.


\bibitem{Med}
  A.~J.~M.~Medved,
  Class.\ Quant.\ Grav.\  19 (2002)  589.

  \bibitem{AWD}
  A.~Ashtekar, J.~Wisniewski, O.~Dreyer,
  Adv.\ Theor.\ Math.\ Phys.\    6 (2003)  507.




\bibitem{JWC}
  Q.~Q.~Jiang, S.~Q.~Wu, X.~Cai,
  Phys.\ Lett.\  B  651 (2007)  58.
\bibitem{mkp3}
  Y.~S.~Myung, Y.~W.~Kim, Y.~J.~Park,
  Phys.\ Rev.\  D   78 (2008)  044020.

\bibitem{Maty}
  J.~Matyjasek,
  Phys.\ Rev.\  D  70 (2004)  047504.

\bibitem{mkp2}
  Y.~S.~Myung, Y.~W.~Kim, Y.~J.~Park,
  Phys.\ Lett.\  B  659 (2008)  832.





\bibitem{BTZ}
  M.~Banados, C.~Teitelboim, J.~Zanelli,
  Phys.\ Rev.\ Lett.\   69 (1992)  1849.

  \bibitem{AO}
  A.~Achucarro, M.~E.~Ortiz,
  Phys.\ Rev.\  D  48 (1993)  3600.



  \bibitem{LMK}
  D.~Louis-Martinez, G.~Kunstatter,
  Phys.\ Rev.\  D  52 (1995)  3494.

\bibitem{GM}
  D.~Grumiller, R.~McNees,
  JHEP  0704 (2007)  074.


\end{thebibliography}
\end{document}